\documentclass[
    ,final            
  ]
  {aipproc}

\pdfoutput=1
\usepackage{amsmath,amssymb,textcomp,units,placeins,graphicx,epsfig}
\layoutstyle{8x11single}

\begin{document}

\title{Axion Dark Matter Searches}


\author{Ian P Stern, \\*
on behalf of the ADMX and ADMX-HF collaborations}{
  address={Department of Physics, Univerisity of Florida, Gainesville, FL 32611-8440}
}

\begin{abstract}
Nearly all astrophysical and cosmological data point convincingly to a large component of cold dark matter in the Universe. The axion particle, first theorized as a solution to the strong charge-parity problem of quantum chromodynamics, has been established as a prominent CDM candidate. Cosmic observation and particle physics experiments have bracketed the unknown mass of the axion between approximately a $\mu$eV and a meV. The Axion Dark Matter eXperiement (ADMX) has successfully completed searches between 1.9 and 3.7 $\mu$eV down to the KSVZ photon-coupling limit. ADMX and the Axion Dark Matter eXperiement High-Frequency (ADMX-HF) will search for axions at weaker coupling and/or higher frequencies within the next few years. Status of the experiments, current research and development, and projected mass-coupling exclusion limits are presented.
\end{abstract}

\maketitle


\section{AXION DARK MATTER}

The Standard Model predicts that the strong and the weak forces violate both parity (P) and charge-parity (CP) symmetries. But the strong force has been found to not violate CP \cite{r1}. This contradiction is known as the Strong CP Problem \cite{r2}. In 1977, Peccei and Quinn theorized a solution by spontaneously breaking a hidden global U(1) symmetry \cite{r3}, often referred to as the Peccei-Quinn (PQ) symmetry. Though not originally intended by Peccei and Quinn, Weinberg and Wilczek independently showed that the spontaneous breaking of the PQ symmetry produces a new pseudo-Nambu-Goldstone boson, dubbed the axion \cite{r4,r5}.

While the axion directly results from the Peccei-Quinn solution to the Strong CP Problem, it has also been established as a prominent dark matter candidate \cite{r6}. The axion naturally meets the requirements of a dark matter candidate with no additional modifications necessary. Axions are cold, non-baryonic particles that possess extremely weak coupling to normal matter, and are dominated by gravitational forces \cite{r7,r8}. Further, the axion is predicted to form a Bose-Einstein condensate, which potentially distinguishes it from other cold dark matter candidates through their phase space structure, supported by astronomical observation \cite{r9}.

The mass of the axion arises from the explicit breaking of PQ symmetry by instanton effects, and is given by 
\begin{equation}
m_a \approx 6\mu\text{eV} \frac{10^{12}\text{GeV}}{f_a}.  \label{eq1}
\end{equation}
$f_a$ is the axion decay constant and is proportional to the vacuum expectation value which breaks PQ symmetry \cite{r2,r10,r11}. While the exact mass of the axion is not known, cosmic observation and particle physics experiments have constrained the particle's mass. The duration of the neutrino burst from SN1987A provided the lower bound on the decay constant of $f_a \gtrsim 10^9$ GeV \cite{r11}. The cosmic energy density argument places the upper limit on the decay constant; if $f_a \gtrsim 10^{12}$ GeV, the axion energy density would be to large, causing the early universe to collapse (overclosure) \cite{r7,r8,r12}. From \eqref{eq1}, the allowable mass of the axion is found to be between approximately $\mu$eV and meV \cite{r11,r13}. 

The axion-photon interaction is dictated by the Lagrangian
\begin{equation}
\mathcal{L}_{a\gamma\gamma} = g_{a\gamma\gamma} a \textbf{\textit{E}} \cdot  \textbf{\textit{B}},  \label{eq2}
\end{equation}
where $a$ is the axion field, and \textbf{\textit{E}} and \textbf{\textit{B}} are the electric and magnetic fields of the two propagating photons, respectively (see Figure \ref{fig1}). The coupling constant, $g_{a\gamma\gamma}$, is proportional to the mass of the axion by 
\begin{equation}
g_{a\gamma\gamma} = \frac{\alpha g_\gamma}{2\pi f_a},  \label{eq3}
\end{equation}
where $\alpha$ is the fine structure constant and $g_\gamma$ is a model-dependent constant of order 1 \cite{r14}. From the above equations and the limits on $f_a$, the axion coupling to electromagnetism can be inferred as extremely weak. The coupling to hadronic matter is even weaker still \cite{r13}. Further, the lifetime of dark matter axions within the allowable mass range is found to be vastly greater than the age of the universe (for $m_a$ = 1 $\mu$eV, $\tau_{\nicefrac{1}{2}}$ $\approx 10^{54}$ s) \cite{r15}. 

Two theoretical models bound the axion-photon coupling constant linearly to the mass by defining $g_\gamma$. KSVZ (Kim-Shifman-Vainshtein-Zakharov) \cite{r16,r17} provides the stronger coupling limit,
\begin{equation}
g_{a\gamma\gamma}^{\text{KSVZ}} \approx 0.38\frac{m_a}{\text{GeV$^2$}},  \label{eq4}
\end{equation}
and DFSZ (Dine-Fischler-Srednicki-Zhitnitskii) \cite{r18,r19} yields the weaker limit,
\begin{equation}
g_{a\gamma\gamma}^{\text{DFSZ}} \approx 0.14\frac{m_a}{\text{GeV$^2$}}.  \label{eq5}
\end{equation}

\section{AXION DARK MATTER DETECTORS}

Because low-mass axions have extremely low decay rates and exceptionally weak interactions with hadronic matter and electromagnetism, they were originally thought to be "invisible" to traditional observational technology \cite{r20}. However, Sikivie showed that the decay of dark matter axions is accelerated within a static magnetic field through the inverse Primakoff effect \cite{r21,r22}.

\begin{figure}[!b]
  \begin{minipage}[b]{.4\linewidth}
    \centering \includegraphics[trim=0 4.2in 0 .62in, clip=true,height=.2\textheight]{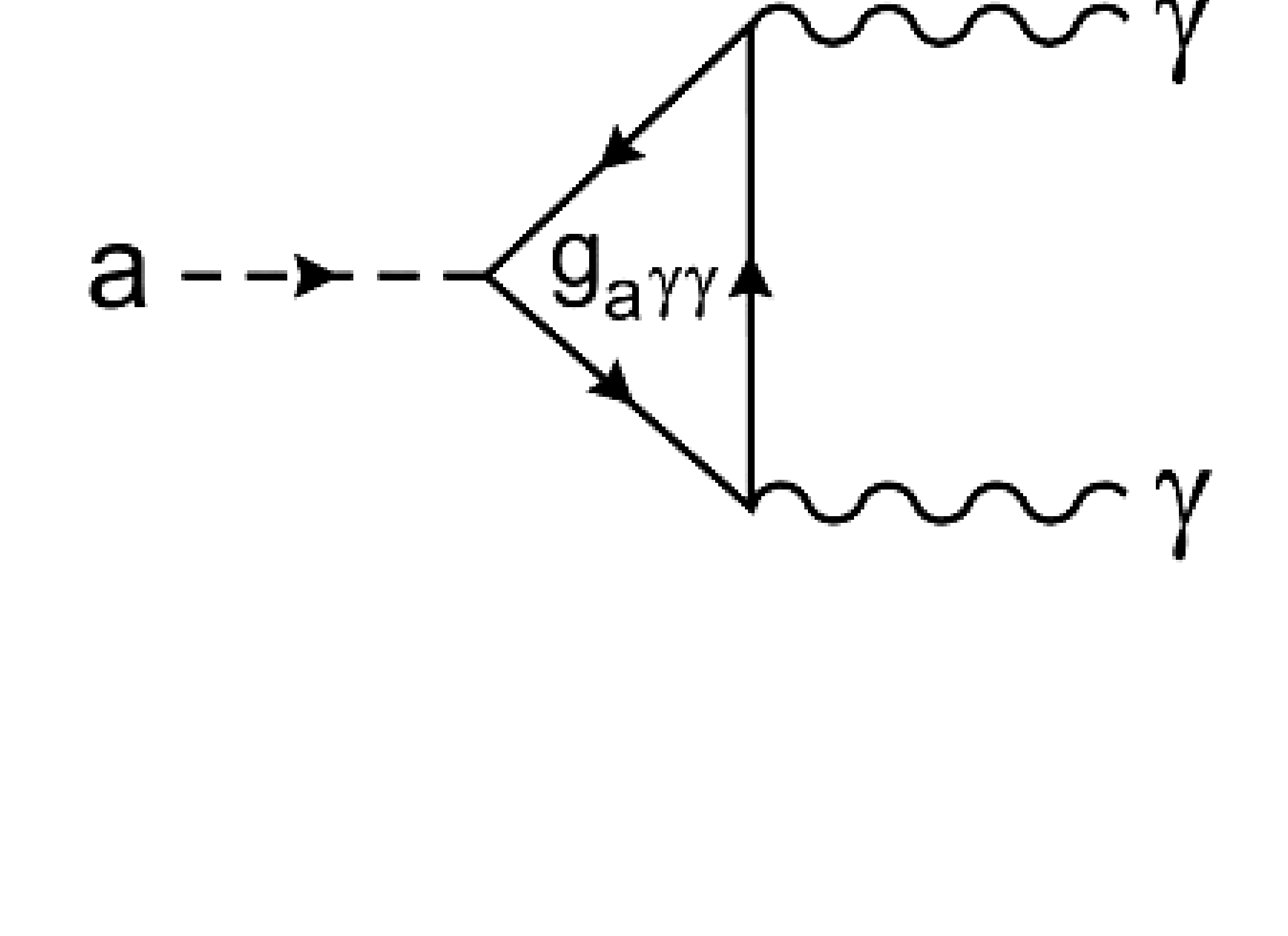}
  \end{minipage}%
  \begin{minipage}[b]{.4\linewidth}
    \centering \includegraphics[trim=0 4.2in 0 .62in, clip=true,height=.2\textheight]{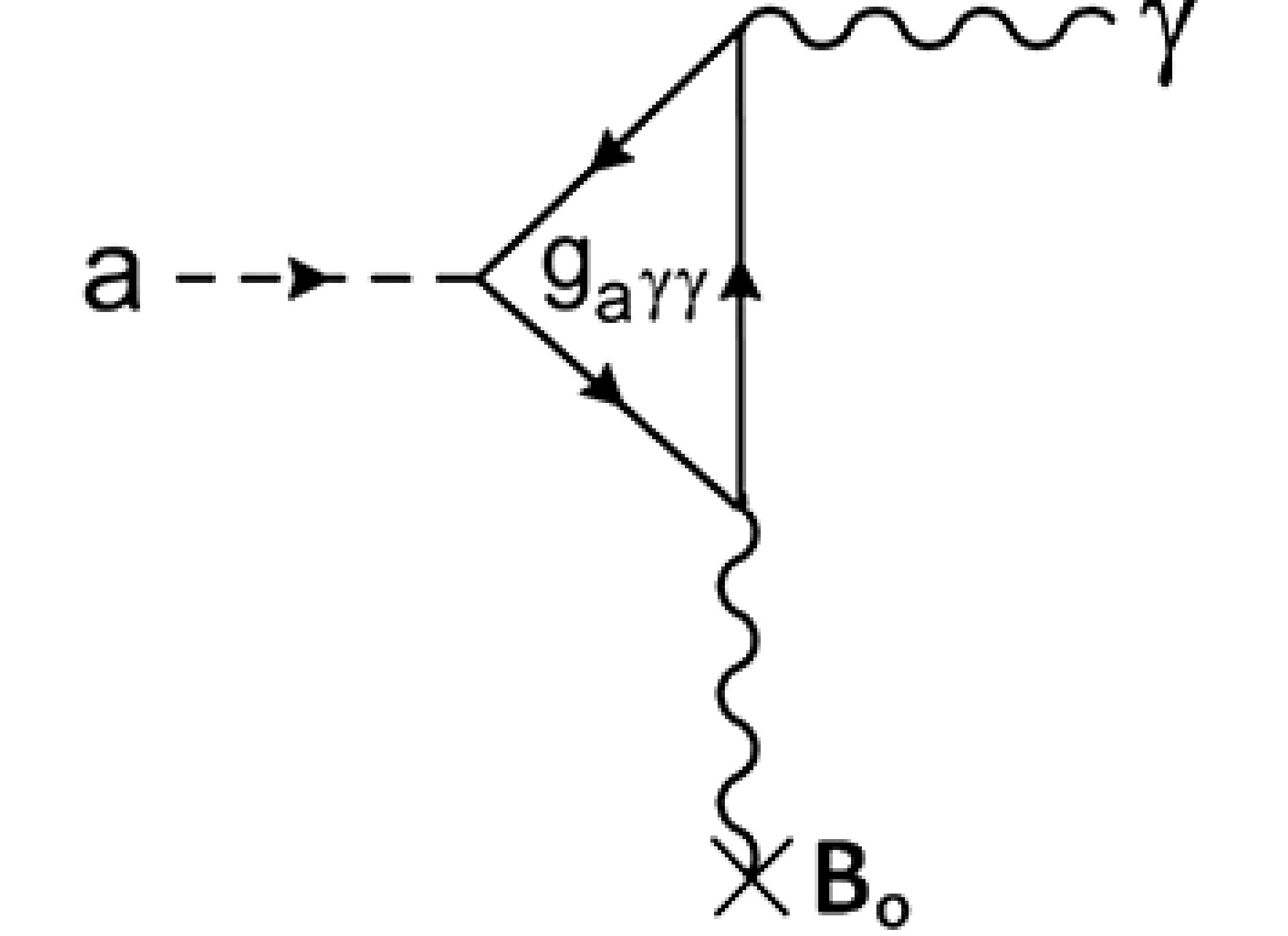}
  \end{minipage}
  \caption{Feynman diagram of axion decay into photons. Left) Conversion in vacuum. Right) Inverse Primakoff effect in a static magnetic field (\textbf{B$_\textbf{o}$}).}
  \label{fig1}
\end{figure}

In a static external magnetic field, one photon is "replaced" by a virtual photon, while the other maintains the energy of the axion, equal to the rest-mass energy ($m_ac^2$) plus the nonrelativistic kinetic energy. \textbf{\textit{B}} in \eqref{eq2} is effectively changed to the static magnetic field, \textbf{\textit{B$_\textbf{o}$}}. Thus, as the magnetic field strength is increased, so does the decay rate of the axion. Figure \ref{fig1} shows the Feynman diagrams for the axion-photon interaction for the two scenarios.

Sikivie proposed an axion detection scheme, based on the Primakoff effect, which used a microwave cavity permeated by a strong magnetic field to resonantly increase the number of photons produced by the decay \cite{r22,r23}. The axion-photon conversion is enhanced when the resonant frequency $f \approx \frac{m_ac^2}{h}$, where $h$ is Planck's constant. There is also a small correction due to the kinetic energy of the axion, but this is tiny ($\frac{{\Delta}E}{E} \approx 10^{-6}$) for cold dark matter. Observing the proper modes at the accurate frequency with significant sensitivity will lead to detection of the axion decay signal. This experiment design is often referred to as a "Sikivie-type" detector.

From \eqref{eq2}, the coupling strength of the axion to the resonant mode is shown to be proportional to $\int d^3\!x\, \textbf{\textit{B$_\textbf{o}$}} \cdot  \textbf{\textit{E$_{\textbf{mnp}}$}}(x)$, where \textbf{\textit{E$_{\textbf{mnp}}$}} is the peak electric field of the mode. The power produced in the cavity is given by
\begin{equation}
P_{mnp} =  g_{a\gamma\gamma}^2 \frac{\rho_a}{m_a} B_o^2 V C_{mnp} Q_L,  \label{eq6}
\end{equation}
where $\rho_a$ is the local energy density of the axion field, $V$ is the volume of the cavity, and $Q_L$ is the loaded quality factor of the cavity (assumed to be less than $Q$ of the axion) \cite{r23}. The subscript on $P_{mnp}$ indicates that the power produced is mode specific. $C_{mnp}$ is the normalized coupling form factor of the axion to a specific mode, defined as
\begin{equation}
C_{mnp} =  \frac{|{\int\limits_V d^3\!x\, \textbf{\textit{B$_\textbf{o}$}} \cdot  \textbf{\textit{E$_{\textbf{mnp}}$}}(x)}|^2}{B_o^2 V{\int\limits_V d^3\!x\, \epsilon(x) |{\textbf{\textit{E$_{\textbf{mnp}}$}}(x)}|^2}}.  \label{eq7}
\end{equation}

\section{AXION DARK MATTER SEARCHES}

\subsubsection{Axion Dark Matter eXperiment (ADMX)}

\begin{figure}
  \includegraphics[trim=.5in .5in .5in 9.12in, clip=true,width=.9\textwidth]{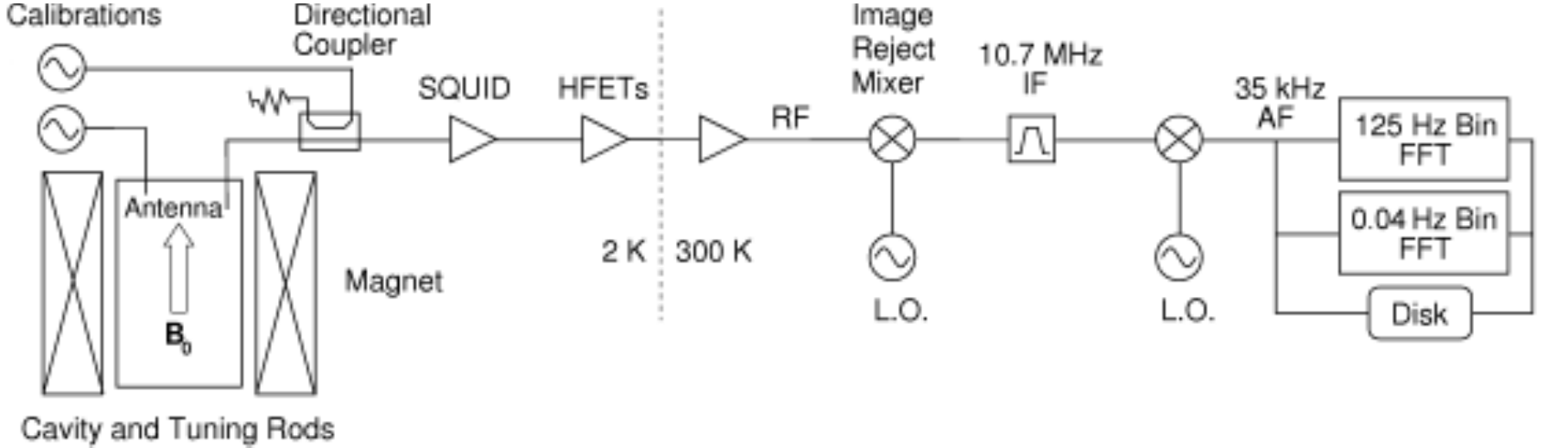}
  \caption{Schematic diagram of the ADMX receiver and electronics. Signals originating in the microwave cavity are amplified by a SQUID and two HFETs (at 2 K), and two post-amplifiers (at 300 K) before being mixed down to audio frequencies and written to disk. The left antenna is a weakly-coupled input antenna used to characterize the resonant modes. }
  \label{fig2}
\end{figure}

Currently located at the University of Washington, the DOE funded ADMX has been the leading contributor to the search for cold dark matter (CDM) axions. Based on Sikivie's proposed scheme, the program uses a large superconducting solenoid, with a nearly homogenous $\sim$7.6 Tesla (T) field, to convert dark matter axions into low energy photons. A cylindrical microwave cavity with a radius ($R$) of 21 cm and a height of 100 cm is used to resonate the decay signal. An output antenna is critically coupled to the cavity to maximize the signal measurement. The output power is increased through a receiver chain containing a SQUID (superconducting quantum interference device) amplifier and two HFET (heterostructure field-effect transistor) amplifiers. Two post-amplifiers and two intermediate-frequency amplifiers provide additional gain at room temperature \cite{r24}. The signal is mixed down into high-resolution and medium-resolution bins and digitally stored. Figure \ref{fig2} illustrates a schematic diagram of the complete receiver chain \cite{r25}. 

The system noise temperature ($T_s$) of the detector, consisting of the physical cavity temperature ($T$) and thermal noise of the detection electronics ($T_n$), had been maintained at approximately 3 K ($T$ = 2 K and $T_n$ = 1 K) using a pumped $^4$He system. ADMX is upgrading the cryogenic system of the experiment in two phases: Phase II and Gen 2. Phase II will install a pumped $^3$He system which will reduce the system noise temperature to 0.6 K. Gen 2 will install a dilution refrigerator which will reduce the system noise temperature to 0.15 K. Because the search rate of the detector is proportional to $T_s^{-2}$, the addition of the pumped $^3$He system and dilution refrigerator will enable ADMX to scan $\sim$25x and $\sim$400x faster, respectively, than previous phases.

Along with the cryogenic system enhancements, ADMX will incorporate a second receiver channel that will search at higher frequencies. The second channel will utilize a separate SQUID amplifier designed to operate above 1 GHz. With the two channels, ADMX will search for axions in the mass ranges of 2.4-3.7 $\mu$eV (580-890 MHz) and 4.9-6.2 $\mu$eV (1.2-1.5 GHz). The program predicts that the planned searches will be sensitive to DFSV coupling limit during Gen 2 operations (see Figure \ref{fig3}).

\subsubsection{Axion Dark Matter eXperiment High-Frequency (ADMX-HF)}

ADMX-HF is a primarily NSF funded axion detection experiment at Yale University that uses a Sikivie-type scheme similar to ADMX, but is designed to search for axions at higher frequencies. The program began in June 2012 and is currently in the fabrication stage. ADMX-HF will use an extremely homogeneous 9 T magnet, and a microwave cavity with a radius of 5 cm and a height of 25 cm. Instead of a SQUID amplifier, the detector will utilize a Josephson Parametric Amplifier (JPA) capable of operating at 4-8 GHz. A dilution refrigerator will maintain the system noise temperature below 0.1 K. 

\begin{figure}
  \includegraphics[trim=.52in 5.57in .48in .50in, clip=true,width=.8\textwidth]{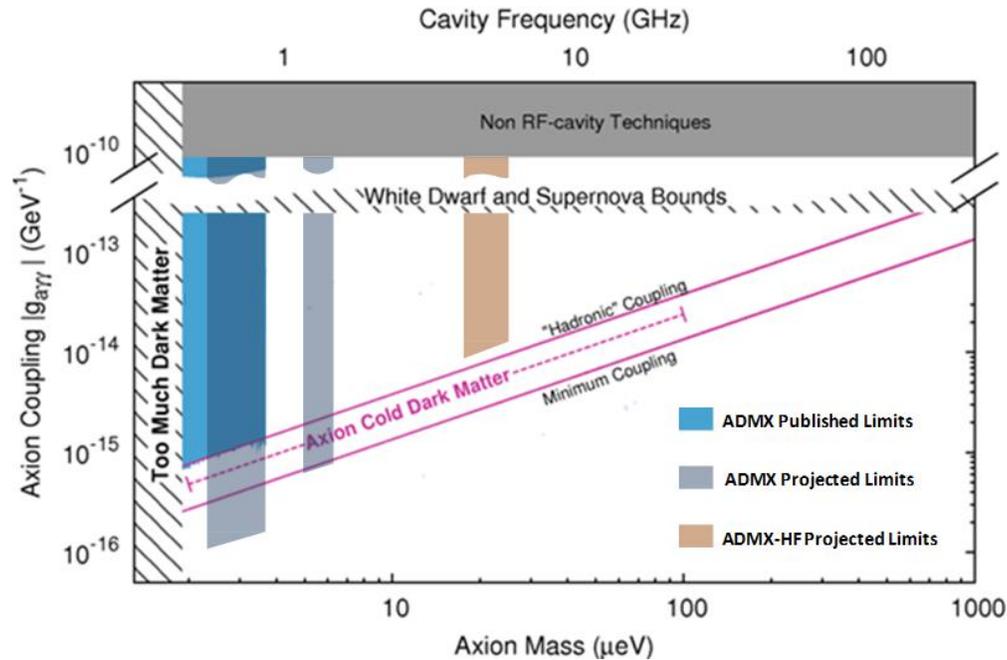}
  \caption{Axion exclusion plot showing published and "near-term" projected sensitivity of ADMX and ADMX-HF programs. The KSVZ limit is labeled "'Hadronic' Coupling" and the DFSZ limit is labeled "Minimum Coupling." Projected limits for ADMX include Phase II and Gen 2 searches. Projected limits for ADMX-HF are for the initial-phase search only.}
  \label{fig3}
\end{figure}

The ADMX-HF detector will be capable of scanning for CDM axions in the range of 19-24 $\mu$eV (4.6-5.9 GHz). However, because the volume of the microwave cavity is significantly smaller than that of ADMX ($\sim$1\% V$_{\text{ADMX}}$), initial searches are not expected to detect axions at the KSVZ coupling limit. Figure \ref{fig3} shows the axion exclusion plot with ADMX published limits and the projected limits of ADMX's and ADMX-HF's near term searches (by 2015). The plot includes the ADMX upgrades for Phase II and Gen 2 and the initial phase for ADMX-HF. 

\subsubsection{Research and Development}

Significant research and development (R\&D) is being conduct by ADMX and ADMX-HF. Since both programs will benefit from most of the technology advancements, much of the R\&D is being performed jointly. The research includes advanced cavity technology, quantum limited electronics, and high-powered uniform magnets.

\begin{figure}
  \begin{minipage}[b]{.427\linewidth}
    \centering \includegraphics[trim=.5in 5.15in .5in .5in, clip=true,height=.2\textheight]{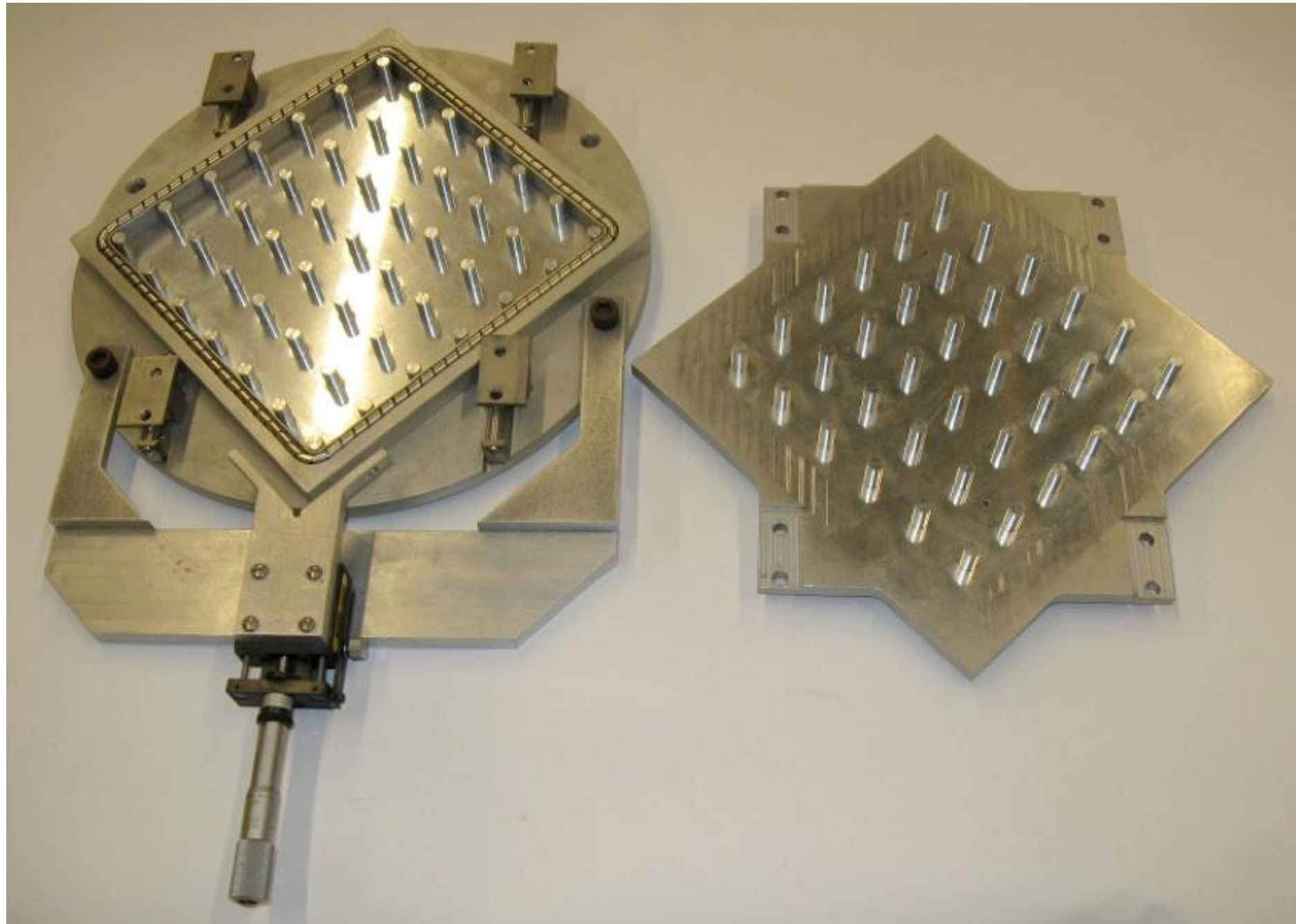}
  \end{minipage}%
  \begin{minipage}[b]{.345\linewidth}
    \centering \includegraphics[trim=.5in 4.9in .5in .5in, clip=true,height=.2\textheight]{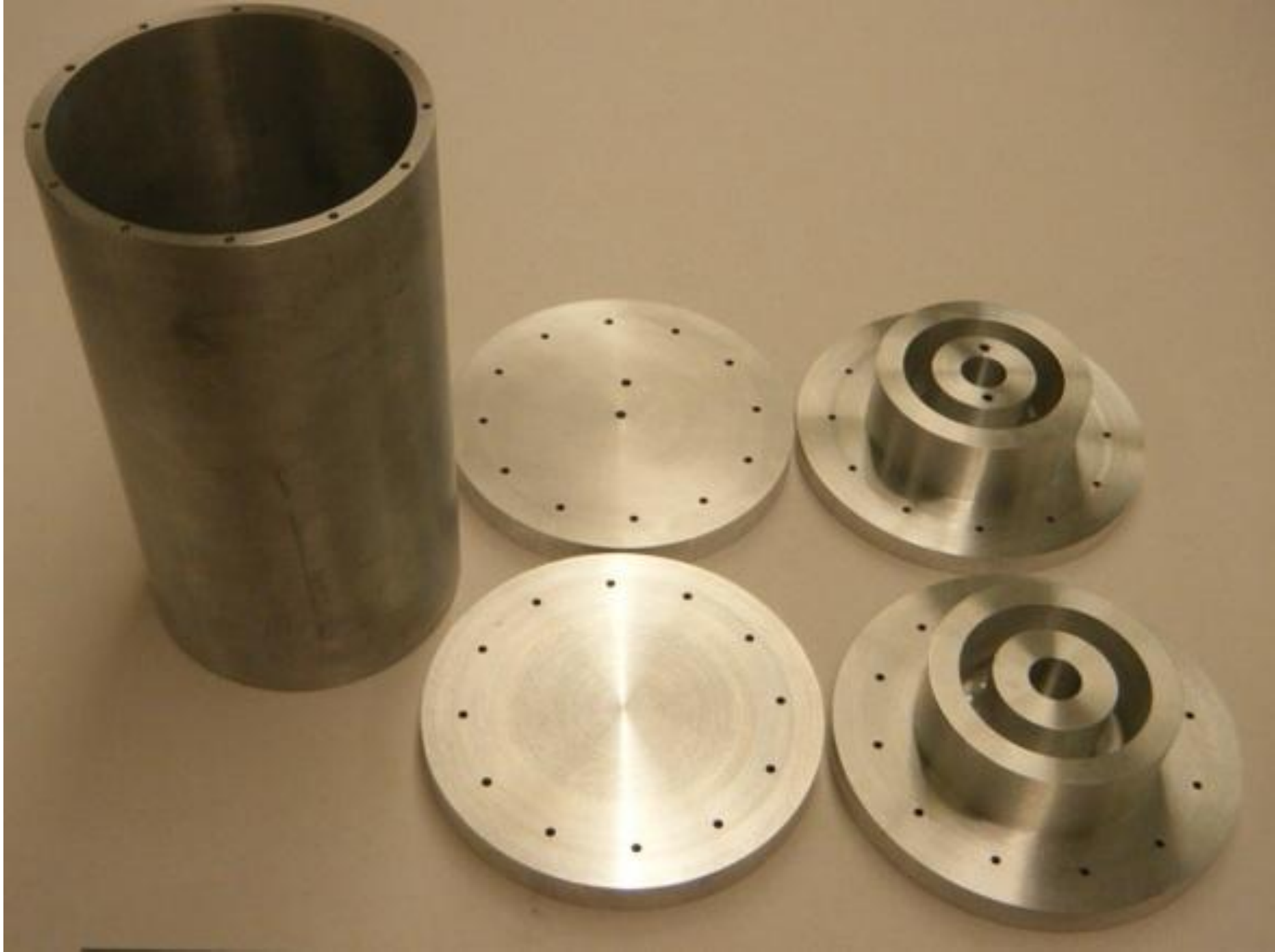}
  \end{minipage}
  \begin{minipage}[b]{.223\linewidth}
    \centering \includegraphics[trim=.5in .5in 3.2in .5in, clip=true,height=.24\textheight]{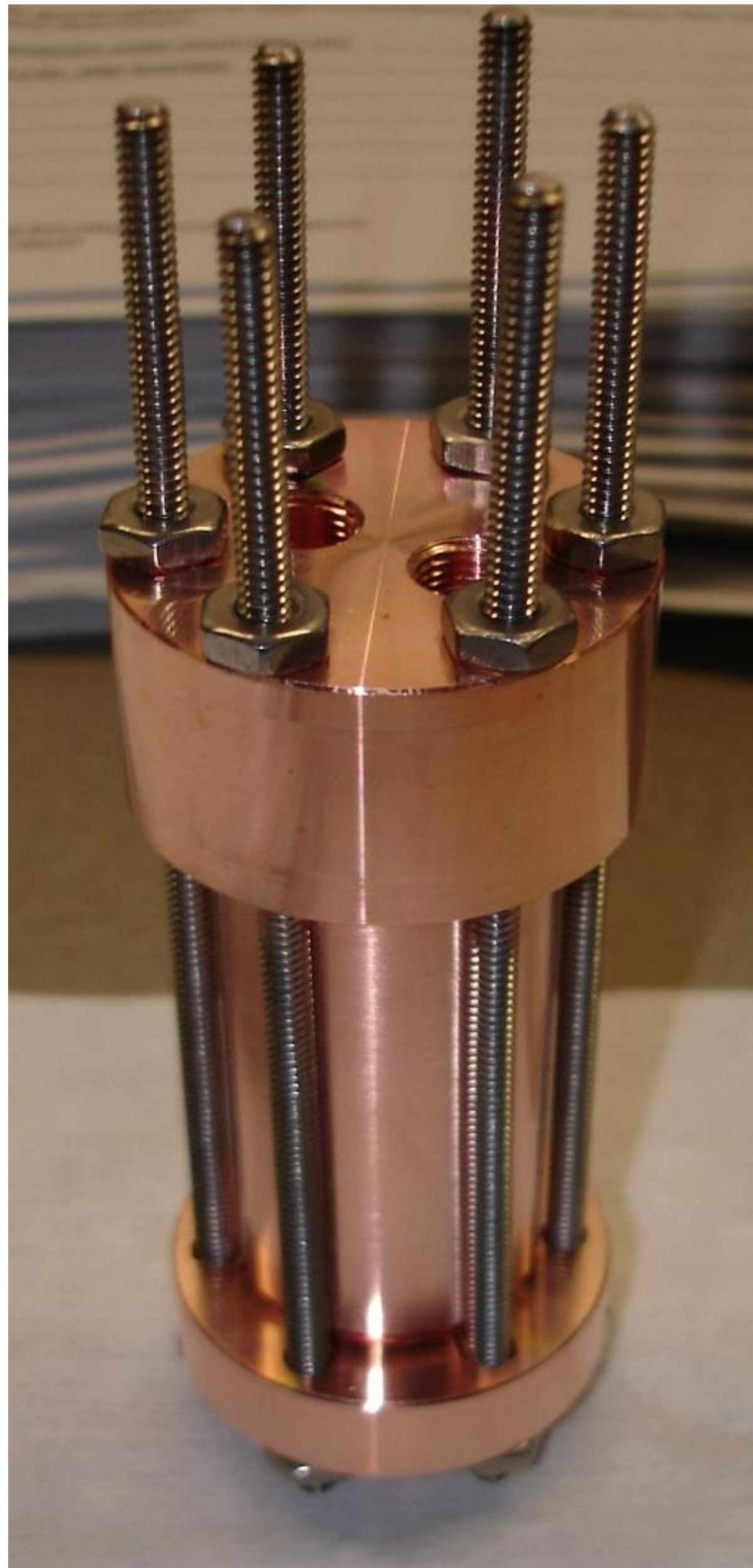}
  \end{minipage}
  \caption{ADMX and ADMX-HF microwave cavity development. Left) Photonic band-gap prototype. Middle) Reactive endcap test cavities. Right) Test fixture for superconducting thin-film coatings in a homogeneous magnetic field.}
  \label{fig4}
\end{figure}

Microwave cavity innovation is critical to the future success of ADMX and ADMX-HF. In order for both programs to extend their search capabilities, new frequency altering and tuning methods need to be developed and tested. ADMX-HF has begun testing of a six post photonic band-gap resonator that increases the search-mode frequencies up to 3 fold over those of an empty cavity. ADMX is investigating photonic band-gaps with higher post densities to obtain even greater frequency increases, as well as reactive endcap designs to decrease the search frequencies. Further, ADMX-HF is conducting a study of superconducting thin-films to produce hybrid cavities for use in a highly-homogeneous magnetic field. The hybrids will have superconductive coating on all surfaces parallel to the magnetic field, greatly reducing the power loss, resulting in an increase in the cavity's quality factor ($Q$) by up to 50 fold. This significant innovation will enable ADMX-HF to conduct searches below the KSVZ coupling limit. Figure \ref{fig4} shows the current stages of several cavity development projects. 

To accommodate higher frequency searches, ADMX and ADMX-HF are advancing SQUID amplifier and JPA capabilities well into the GHz range. ADMX-HF has successfully tested JPAs up to 8 GHz, and plans to test devices beyond 10 GHz in the near future. ADMX is currently testing a SQUID amplifier up to 1.5 GHz, and is researching methods to advance the usable frequencies above 2 GHz. Both programs are approaching the quantum limits for thermal noise of the devices.

\begin{figure}
  \includegraphics[trim=.54in 5.59in .47in .53in, clip=true,width=.8\textwidth]{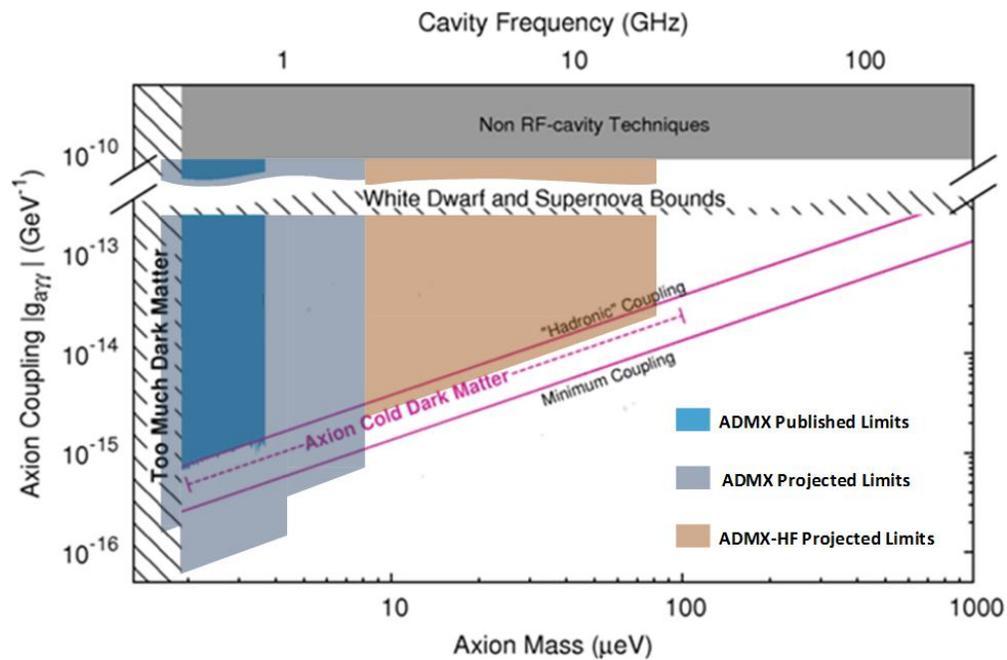}
  \caption{Axion exclusion plot showing published and "long-term" projected sensitivity of ADMX and ADMX-HF programs. The KSVZ limit is labeled "'Hadronic' Coupling" and the DFSZ limit is labeled "Minimum Coupling." Projected limits for ADMX include anticipated results of current research and development for ADMX and ADMX-HF, as well as Phase II and Gen 2 searches.}
  \label{fig5}
\end{figure}

ADMX-HF has commissioned a 9 T superconducting solenoid with a radial component of the magnetic field that is less than 0.005 T throughout the usable volume. The magnet can accommodate a 2 liter microwave cavity and is designed to prevent flux quanta penetrating the superconducting coating of hybrid cavities. ADMX is also investigating upgrading the detector magnet to increase the field strength and maintain a 0.005 T radial component. Doubling the magnetic field strength will quadruple the signal power of the experiment. Figure \ref{fig5} shows the exclusion plot with ADMX published limits and the projected limits of ADMX's and ADMX-HF's far term searches (< 10 years). The plot includes the anticipated detector capability enhancements of the current R\&D projects, as well as the ADMX upgrades for Phase II and Gen 2 (see Figure \ref{fig3}). 

\section{CONCLUSION}

The "invisible" axion has been shown to be a strong cold dark matter candidate within the mass range of approximately $\mu$eV and meV. ADMX is the only experiment to exclude CDM axions at the KSVZ coupling limits. ADMX and ADMX-HF will soon extend on the success of the Sikivie-type axion detectors by searching at weaker couplings and/or higher masses within the next few years. Both programs are conducting long-term R\&D projects that will greatly expand their capabilities, which is predicted to result in completing searches over approximately 50\% of the allowable mass-coupling range of axion dark matter within the next decade. 


\begin{theacknowledgments}
Funding for ADMX was provided by the Department of Energy Grants DE-FG02-97ER41029, DE-FG02-96ER40956, DEAC52-07NA27344, and DE-AC03-76SF00098, and the Livermore LDRD program. Funding for ADMX-HF was provided by the National Science Foundation Grant 1067242.
\end{theacknowledgments}



\bibliographystyle{aipproc}   

\bibliography{references}

\IfFileExists{\jobname.bbl}{}
 {\typeout{}
  \typeout{******************************************}
  \typeout{** Please run "bibtex \jobname" to optain}
  \typeout{** the bibliography and then re-run LaTeX}
  \typeout{** twice to fix the references!}
  \typeout{******************************************}
  \typeout{}
 }

\end{document}